\documentclass[twocolumn,showkeys,aps,prb,showpacs]{revtex4-1}
\UseRawInputEncoding
\usepackage{graphicx}
\usepackage[CJKbookmarks,dvipdfm,colorlinks,linkcolor=blue,citecolor=blue]{hyperref}

\begin{document}

\title{Correlation-enhanced spin-orbit coupling and quantum anomalous Hall insulator with large band gap and stable ferromagnetism in monolayer  $\mathrm{Fe_2Br_2}$}

\author{San-Dong Guo$^{1}$, Yu-Tong Zhu$^{1}$  and  Bang-Gui Liu$^{2,3}$}
\affiliation{$^1$School of Electronic Engineering, Xi'an University of Posts and Telecommunications, Xi'an 710121, China}
\affiliation{$^2$ Beijing National Laboratory for Condensed Matter Physics, Institute of Physics, Chinese Academy of Sciences, Beijing 100190, People's Republic of China}
\affiliation{$^3$School of Physical Sciences, University of Chinese Academy of Sciences, Beijing 100190, People's Republic of China}
\begin{abstract}
 Nontrivial band topology  combined with  magnetic ordering can  produce quantum anomalous Hall  insulator (QAHI), which  may lead
to advances in device concepts. Here, through first-principles calculations,  stable monolayer  $\mathrm{Fe_2Br_2}$ is predicted as a room-temperature large-gap high-Chern-number QAHI by using  generalized gradient approximation plus $U$ (GGA+$U$) approach. The large gap is due to  correlation-enhanced  spin-orbit coupling (SOC) effect of Fe atoms, which equates with artificially increasing the strength of SOC  without electronic correlation. Out-of-plane magnetic anisotropy is very key to produce quantum anomalous Hall (QAH) state because in-plane magneitization will destroy nontrivial band topology. In the absence of SOC, $\mathrm{Fe_2Br_2}$ is a half Dirac semimetal state protected by mirror symmetry, and the electronic correlation along with SOC effect creates QAH state with a
sizable gap and two chiral edge modes.  It is found that the QAH state is robust against biaxial strain ($a/a_0$: 0.96 to 1.04) in monolayer  $\mathrm{Fe_2Br_2}$  with stable ferromagnetic (FM) ordering and out-of-plane magnetic anisotropy. Calculated results show that Curie temperature
is sensitive to correlation strength and strain. The  reduced correlation and compressive strain are  in favour of high Curie temperature.
These analysis and results  can be readily extended to other monolayer $\mathrm{Fe_2XY}$ (X/Y=Cl, Br and I), which possesses
the same Fe-dominated  low-energy states with a  $\mathrm{Fe_2Br_2}$ monolayer. These findings open
new opportunities to   design new high-temperature topological quantum devices.

\end{abstract}
\keywords{Correlation, Spin-orbit coupling, Magnetic ordering, Band topology}

\maketitle

\section{Introduction}
The quantum anomalous Hall effect (QAHE) is characterized by the nonzero Chern numbers with
the quantized Hall conductance\cite{h0}, which can be used to design
low-power-consumption spintronic devices  due to the existence of
dissipationless chiral edge states under zero magnetic field. The QAH state also provides a fertile
playground to explore topological magnetoelectric effects, Majorana
fermions and quantum computation\cite{h1,h2,h3}. However,  the experimental observation of  QAHE  is rare, and its ultralow working temperature limits  exploring these emergent quantum physics and promising applications.
The  QAHE is firstly observed in Cr-doped  $\mathrm{(Bi, Sb)_2Te_3}$ thin films at a very low
temperature of 30 mK\cite{h4}. In quick succession, the QAHE is realized in V-doped
and Cr-and-V co-doped $\mathrm{(Bi, Sb)_2Te_3}$ thin film at about 25 and 300 mK\cite{h5,h6}.
 Recently, the intrinsic QAHE has
been observed in the van der Waals (vdW) layered material $\mathrm{MnBi_2Te_4}$\cite{h7,h8}, and a high-Chern-number QAHE has also been
obtained in a 10-layer $\mathrm{MnBi_2Te_4}$ device at about 13 K\cite{h9}. So, seeking high-temperature QAHI with large gap is a compelling problem of condensed matter physics and materials science.

\begin{figure}
  \includegraphics[width=8.0cm]{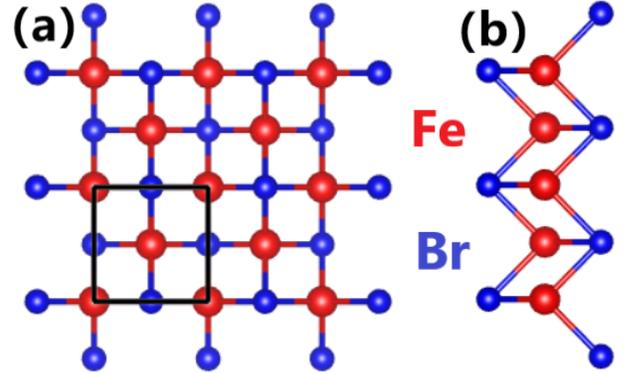}
  \caption{(Color online) The  crystal structure of  monolayer $\mathrm{Fe_2Br_2}$: (a) top view and (b) side view.  The  primitive cell is marked by  black frame.}\label{t0}
\end{figure}

\begin{figure}
  \includegraphics[width=8cm]{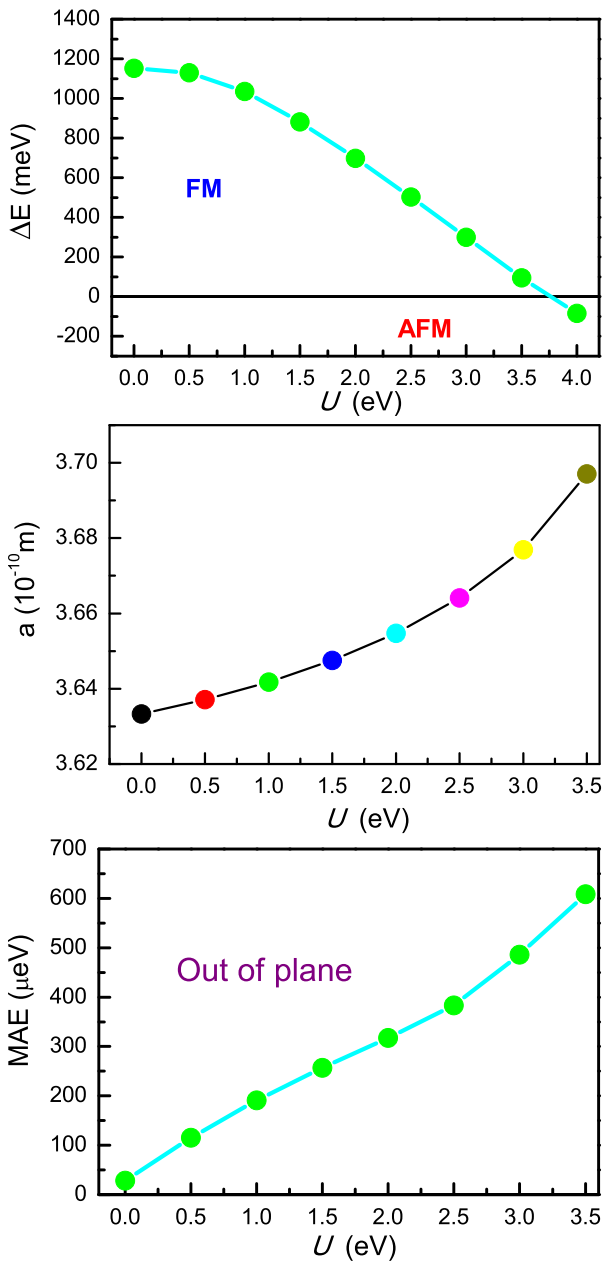}
\caption{(Color online) For monolayer $\mathrm{Fe_2Br_2}$, the energy difference between AFM and FM (Top), lattice constants (Middle) and MAE (Bottom)  as a function of  $U$. }\label{u-eam}
\end{figure}

However, searching for high-temperature QAHI with large nontrivial band gap is challenged. The ferromagnetism prefers metallic systems composed of light 3$d$ elements, while topological insulator (TI) favors heavy elements for
achieving strong SOC effects. Recently,
a robust QAHI  $\mathrm{Fe_2I_2}$ monolayer with centrosymmetry is predicted with  a large nontrivial band gap of
301 meV and a high Curie temperature of about 400 K\cite{fe}.  In  $\mathrm{Li_2Fe_2X_2}$ (X=S, Se and Te) monolayers (Lithium-decorated
iron-based superconductor monolayer materials), the high-temperature large-gap  QAHIs are also achieved\cite{fe1}.
To realize the combination of
piezoelectricity with topological properties, monolayer  $\mathrm{Fe_2IX}$ (X=Cl and Br), $\mathrm{FeI_{1-x}Br_{x}}$ (x=0.25 and 0.75) and  $\mathrm{Li_2Fe_2SSe}$ are predicted\cite{fe2,fe3,fe4}, namely two-dimensional
(2D) piezoelectric quantum anomalous Hall insulator (PQAHI), which provides possibility to use  piezotronic effect to control QAHE. All these 2D materials share the same Fe layer structure and Fe-dominated  low-energy states, and they all are room-temperature large-gap high-Chern-number QAHIs. The SOC  is generally considered to be
appreciable only in heavy elements, and the SOC effect should be very small in light Fe element. However, these 2D materials are predicted to large-gap QAHIs. In ref.\cite{fe1},  a large QAH gap is attributed to  the enlarged effective
SOC strength of $d$ orbitals  by bonding with
heavy elements,  and then the SOC effects are significantly
enhanced near Dirac cones.

To clarify this question,   we take  monolayer  $\mathrm{Fe_2Br_2}$  as a concrete example to study the correlation effects on effective SOC strength.
Why do we choose  monolayer  $\mathrm{Fe_2Br_2}$? Firstly, monolayer  $\mathrm{Fe_2Br_2}$  has not been studied in detail, including its stability and electronic structures.  Secondly, for $\mathrm{Fe_2Br_2}$, the Fe elements  bond with
relatively light elements Br compared to I elements in $\mathrm{Fe_2I_2}$, and it has  a simpler structure than $\mathrm{Li_2Fe_2S_2}$.
In this work, it is found that the  correlation-enhanced  SOC effect of Fe atoms induces large nontrivial band gap in  monolayer  $\mathrm{Fe_2Br_2}$, and the effective enlarged  SOC strength of Fe-3$d$ orbitals is not due to bonding with
heavy elements Br. Calculated results show out-of-plane magnetic anisotropy is necessary to produce QAH state. So, the magnetic anisotropy direction must be determined to achieve QAHE in these 2D materials. Taking classic $U$=2.5 eV\cite{fe,fe5} as an example, $\mathrm{Fe_2Br_2}$ is proved to be  dynamically, mechanically  and thermally stable.  The high Chern number (C=2)  is
firmly confirmed by Berry curvatures and
chiral edge states.  In $\mathrm{Fe_2Br_2}$, the QAH state, FM ordering and out-of-plane magnetic anisotropy are robust against biaxial strain. It is found that reduced correlation and compressive strain make for high Curie temperature. Our works provide basis for understanding large nontrivial band gap in monolayer $\mathrm{Fe_2XY}$ (X/Y=Cl, Br and I) and $\mathrm{Li_2Fe_2XY}$ (X/Y=S, Se and Te).

The rest of the paper is organized as follows. In the next
section, we shall give our computational details and methods.
 In  the next few sections,  we shall present electronic correlation effects on electronic structures along with the case at $U$=2.5 eV,  strain influence on topological properties,  and Curie temperatures of monolayer  $\mathrm{Fe_2Br_2}$.  Finally, we shall give our discussion and conclusion.

\begin{figure}
   \includegraphics[width=7.0cm]{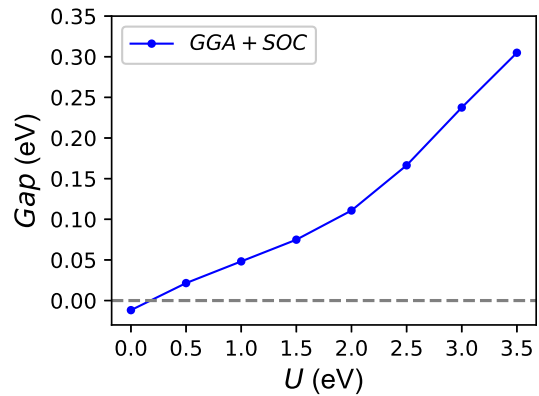}
  \caption{(Color online) For monolayer $\mathrm{Fe_2Br_2}$, the gaps as a function of $U$  using  GGA+SOC.}\label{gap}
\end{figure}

\begin{figure*}
  \includegraphics[width=16cm]{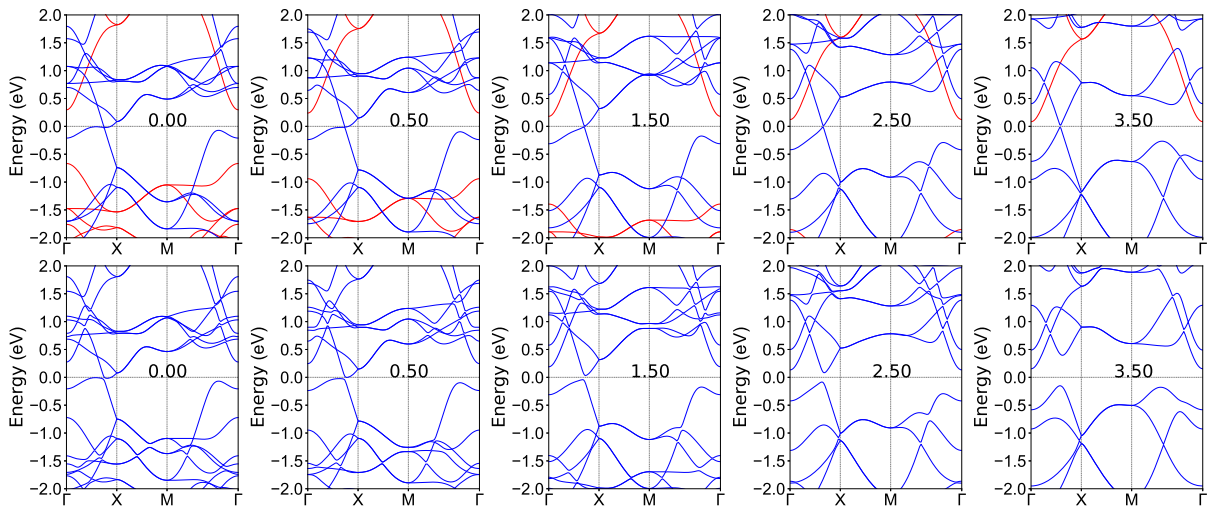}
  \caption{(Color online)For monolayer $\mathrm{Fe_2Br_2}$, the  energy band structures  without SOC (Top) and with SOC (Bottom) as a function of $U$. The red (blue) lines represent the band structure in the spin-up (spin-down) direction without SOC.  }\label{band}
\end{figure*}
\begin{figure}
   \includegraphics[width=8.0cm]{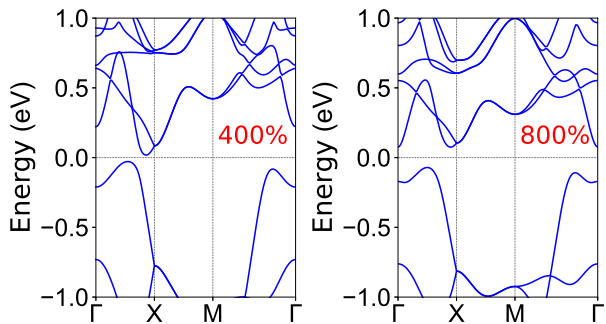}
  \caption{(Color online) The energy band structures of $\mathrm{Fe_2Br_2}$  with  enhanced SOC strength to 400\% and 800\% at $U$=0.00 eV.}\label{band-400}
\end{figure}

\begin{figure*}
   \includegraphics[width=12.0cm]{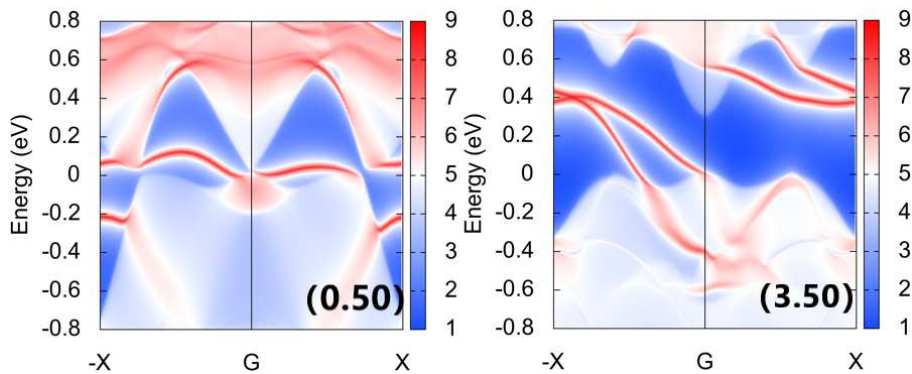}
  \caption{(Color online) Topological
edge states of $\mathrm{Fe_2Br_2}$  calculated  along the (100) direction at $U$=0.50 eV and $U$=3.50 eV.}\label{s-1}
\end{figure*}

\begin{figure*}
  \includegraphics[width=16cm]{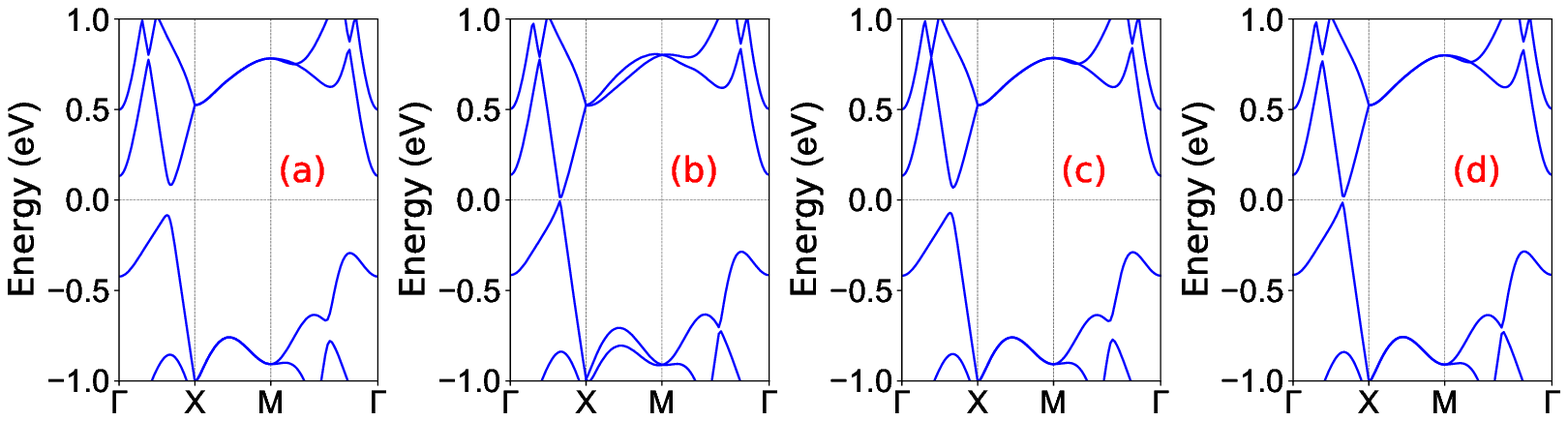}
  \caption{(Color online) For monolayer $\mathrm{Fe_2Br_2}$, the  energy band structures: (a) full SOC with out-of-plane magnetization; (b)full SOC with in-plane magnetization; (c) only considering SOC of Fe with out-of-plane magnetization; (d) only considering SOC of Br with out-of-plane magnetization. }\label{band-2.5}
\end{figure*}

\section{Computational detail}
Within density functional theory (DFT)\cite{1}, the first-principles calculations are performed  by using the Vienna ab initio
simulation package (VASP)\cite{pv1,pv2,pv3} with plane-wave basis set. The projector augmented
wave (PAW) method is adopted in conjugation with a GGA of  Perdew, Burke and
Ernzerhof (PBE-GGA) as exchange-correlation functional\cite{pbe}. A plane-wave cutoff of 500 eV is used to perform geometric optimization and electronic properties
calculations of monolayer $\mathrm{Fe_2Br_2}$.  The energy convergence criterion is set  for $10^{-8}$ eV, and the residual force is less than 0.0001 $\mathrm{eV.{\AA}^{-1}}$. The Brillouin zone (BZ) integration is
carried out with 18$\times$18$\times$1 k-point sampling. To avoid interactions
between two neighboring images, the vacuum
region along the $z$ direction is set to be larger than 16 $\mathrm{{\AA}}$. The electronic correlation of Fe atoms is considered within
the GGA+$U$ scheme by the
rotationally invariant approach proposed by Dudarev et al\cite{u}. The SOC is included self-consistently in the calculations  to investigate
magnetic anisotropy and topological properties.

The elastic stiffness tensor  $C_{ij}$   is calculated by using strain-stress relationship (SSR) with GGA, and
the 2D elastic coefficients $C^{2D}_{ij}$
have been renormalized by   $C^{2D}_{ij}$=$L_z$$C^{3D}_{ij}$ with  $L_z$ being the length of unit cell along $z$ direction.
For phonon spectrum calculation, the Phonopy code\cite{pv5} is used within finite displacement method.  The interatomic force constants (IFCs) with a 5$\times$5$\times$1 supercell are calculated to attain
phonon dispersions. For ab initio
molecular dynamics (AIMD) simulation, the calculation is  carried out
with a 4$\times$4$\times$1 supercell for more than
8000 fs with a time step of 1 fs  by
using canonical ensemble. WannierTools code is used to perform surface state and Berry curvature calculations, based on the tight-binding Hamiltonians constructed from maximally localized Wannier functions by Wannier90 code\cite{w1,w2}. The Curie temperature is estimated by  Monte Carlo (MC) simulation with a  40$\times$40 supercell and  $10^7$ loops, as implemented in Mcsolver code\cite{mc}.

\section{Electronic correlation effects}
As shown in \autoref{t0}, the unit cell of $\mathrm{Fe_2Br_2}$ contains four atoms with two co-planar Fe atoms being sandwiched between two layers of Br atoms. The  $\mathrm{Fe_2Br_2}$ possesses  $P4/nmm$ space group (No. 129) with centrosymmetry.  The key space-group symmetry operations of $\mathrm{Fe_2Br_2}$ include
space inversion $P$, $C_4$ rotation, $M_x$ ($M_y$) mirror and
glide mirror $G_z= \{M_z|\frac{1}{2},\frac{1}{2},0\}$, which is different from monolayer  $\mathrm{Fe_2IBr}$  with missing $P$ and glide mirror $G_z$\cite{fe2}. So, monolayer  $\mathrm{Fe_2IBr}$  possesses piezoelectricity, but $\mathrm{Fe_2Br_2}$ is not piezoelectric.
it is noteworthy  that magnetic orientation can influence these symmetries within SOC.

In the presence of the electron correlation, the SOC effect of  3$d$  systems with special orbital symmetry and electron occupation is found to be more prominent\cite{h10,h11,h12,h13}. Here, we optimize
lattice constants $a$  with varied $U$ (0-4 eV), and then  determine magnetic ground state. The  energy differences between antiferromagnetic (AFM) and FM ordering vs $U$ are plotted in \autoref{u-eam}. With increasing $U$, the ground state of  $\mathrm{Fe_2Br_2}$ changes from FM ordering to AFM one, and the critical $U$ value is about 3.75 eV. It is found that increasing $U$ weakens FM interaction, giving rise to important influence on Curie temperature of  $\mathrm{Fe_2Br_2}$. The $a$ as a function of $U$ with FM ground state is shown in \autoref{u-eam}, and $a$ increases with increasing $U$. Similar result can be found in FeClF monolayer\cite{h12}.
The different magnetic orientation can  affect the symmetry of 2D systems, and then produce important influence on
their valley and topological properties\cite{h10,h12}. The magnetic anisotropy can be described by magnetic anisotropy energy (MAE), which plays
a very important role to determine  thermal stability of magnetic
ordering.  The MAE can be calculated by $E_{MAE}$ = $E_x$-$E_z$,  where $E_{x/z}$ is the energy per Fe atom when the magnetization is along the $x/z$ direction.  In \autoref{u-eam}, we plot MAE as a function of $U$, which  favors an out-of-plane FM state in considered $U$ range. The out-of-plane magnetic anisotropy plays a crucial role to produce QAH state, as we shall see in a while. It is found that increasing $U$ makes for out-of-plane  one.
This is different from one of FeClF monolayer\cite{h12}, where increasing $U$ changes its magnetic anisotropy from out-of-plane to in-plane.

Next, we prove that  electron correlation can dramatically enhance the SOC effect in $\mathrm{Fe_2Br_2}$ monolayer. The energy band structures with varied $U$ are calculated by using GGA and GGA+SOC. The  GGA+SOC gaps as a function of $U$ are shown in \autoref{gap}, and both GGA and GGA+SOC energy bands  at representative $U$ values are plotted in \autoref{band}.  Ignoring electron correlation  by setting $U$=0.00 eV, the SOC induce a very small splitting along $\Gamma$-X line near the Fermi level.
However, a metallic state is produced, because both conduction and valence bands slightly cross the Fermi level. Once the correlation effect is included  self-consistently, the SOC-induced gap is enhanced
dramatically, reaching 305 meV  ($U$=3.5 eV). To further confirm correlation-enhanced SOC,  we artificially increase the strength of SOC,  and
realize a large energy gap in  $\mathrm{Fe_2Br_2}$ without  electron correlation ($U$=0.00 eV).  The
SOC strength is improved to  400\%/800\% of the normal one, and the corresponding energy bands are plotted in \autoref{band-400}. It is clearly seen that a gap of 47/146 meV is induced with   enhanced SOC strength to 400\%/800\%. The gap with  enhanced SOC strength to 800\% at $U$=0.00 eV is close to one (166 meV) with the normal SOC strength at $U$=2.5 eV.
These show that correlation indeed can  trigger enhanced SOC effect, and then produce large-gap QAHI.
\begin{figure*}
   \includegraphics[width=13.0cm]{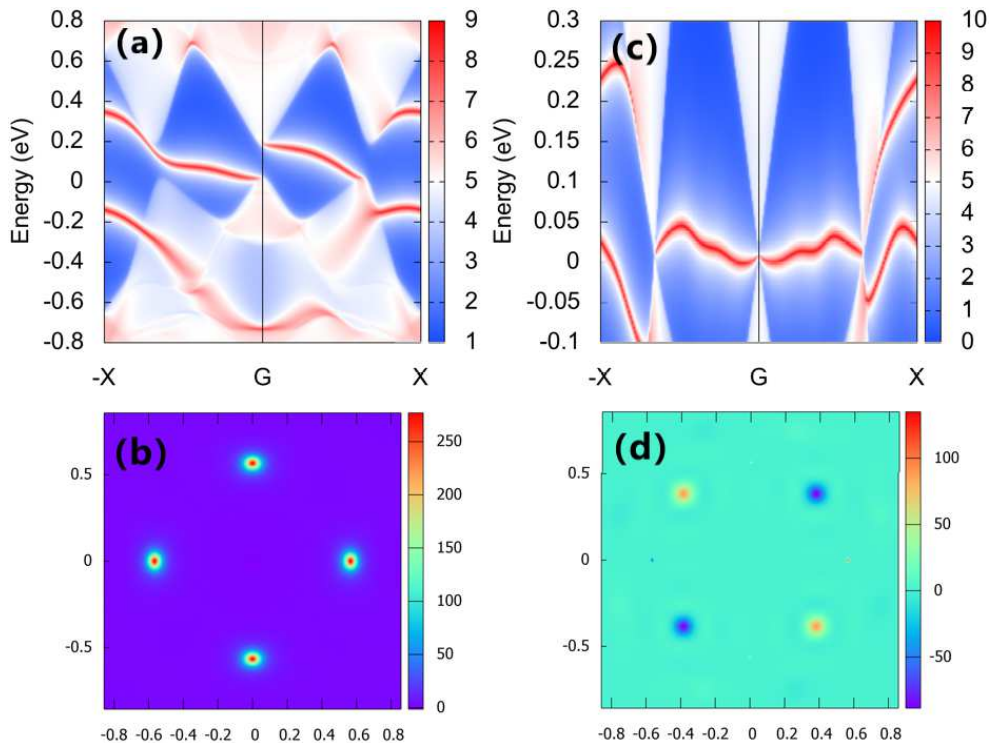}
  \caption{(Color online)For monolayer $\mathrm{Fe_2Br_2}$, topological
edge states (a,c) and  distribution of Berry curvature  (b,d)  at $U$=2.50 eV with out-of-plane (a.b) and in-plane (c,d) magnetization.}\label{s-2}
\end{figure*}

Within GGA, in small $U$ range, distorted  Dirac cones can be observed with the valence
and conduction energy bands of the minority-spin touching   the Fermi
level. With increasing $U$,  there are  Dirac cones with linear
band dispersion, for example  $U$=2.50 eV.
For Dirac states, the presence of SOC  triggers gap
opening at the touching points, which means that $\mathrm{Fe_2Br_2}$ should be  a QAHI.
 By  checking  the chiral edge modes (see \autoref{s-1}),
we therefore confirm  that monolayer $\mathrm{Fe_2Br_2}$ is a potential QAHI in considered $U$ range. It is clearly seen that two chiral edge states  does exist, which connect  the conduction bands and
valence  bands.  This means that the Chern number of $\mathrm{Fe_2Br_2}$ is equal to two ($C$=2). The $U$ dependence of electronic structures in $\mathrm{Fe_2Br_2}$ is different from those in monolayer $\mathrm{VSi_2P_4}$, $\mathrm{FeCl_2}$ and FeClF\cite{h10,h11,h12}, and they undergo a rich topological
phase transition with  QAH phase only existing in certain $U$ range.

\section{The case at $U$=2.5 eV}
We adopt typical $U$=2.5 eV\cite{fe,fe5} as  a concrete example to detailedly investigate  the physical properties of $\mathrm{Fe_2Br_2}$. To confirm the stability of  $\mathrm{Fe_2Br_2}$, phonon dispersion,  AIMD
simulation  and elastic constants $C_{ij}$  are calculated by using GGA. As shown in FIG.1 of electronic supplementary information (ESI), the phonon spectra of $\mathrm{Fe_2Br_2}$ shows no imaginary frequency,  indicating its dynamical stability. By  AIMD simulation, the temperature and total energy fluctuations of $\mathrm{Fe_2Br_2}$ as a function of  simulation time along with final structures  after 8 ps are plotted in FIG.2 of ESI at 300 K. Neither structure reconstruction nor bond breaking with small  temperature and total energy
 fluctuations  confirm its thermodynamical stability at room temperature.
We use elastic constants to prove  mechanical stability of monolayer $\mathrm{Fe_2Br_2}$. Using Voigt notation, the 2D  elastic tensor with space group $P4/nmm$  can be expressed as:
\begin{equation}\label{pe1-4}
   C=\left(
    \begin{array}{ccc}
      C_{11} & C_{12} & 0 \\
     C_{12} & C_{11} &0 \\
      0 & 0 & C_{66} \\
    \end{array}
  \right)
\end{equation}
The three independent elastic constants $C_{11}$, $C_{12}$ and  $C_{66}$ are 48.69 $\mathrm{Nm^{-1}}$, 22.78 $\mathrm{Nm^{-1}}$ and 29.01 $\mathrm{Nm^{-1}}$,  which  satisfy the  Born  criteria of mechanical stability:
$C_{11}>0$, $C_{66}>0$, $C_{11}-C_{12}>0$,  confirming  its mechanical stability.

Due to $P4/nmm$ symmetry,      $\mathrm{Fe_2Br_2}$ is  mechanically anisotropic.
The direction-dependent in-plane Young's moduli $C_{2D}(\theta)$ and
Poisson's ratios $\nu_{2D}(\theta)$  can be attained from the calculated $C_{ij}$ by using these expressions\cite{ela,ela1}:
 \begin{equation}\label{pe1-4-1}
  C_{2D}(\theta)=\frac{C_{11}C_{22}-C_{12}^2}{C_{11}m^4+C_{22}n^4+(B-2C_{12})m^2n^2}
\end{equation}
 \begin{equation}\label{pe1-4-2}
  \nu_{2D}(\theta)=\frac{(C_{11}+C_{22}-B)m^2n^2-C_{12}(m^4+n^4)}{C_{11}m^4+C_{22}n^4+(B-2C_{12})m^2n^2}
\end{equation}
in which  $m=sin(\theta)$, $n=cos(\theta)$ and $B=(C_{11}C_{22}-C_{12}^2)/C_{66}$. The $\theta$ is the angle of the direction with
 the $x$ direction  as $0^{\circ}$ and $y$ direction as $90^{\circ}$.
The direction-dependent Young's moduli $C_{2D}(\theta)$ and
Poisson's ratios $\nu_{2D}(\theta)$   are plotted in FIG.3 of ESI.
Due to cubic symmetry,  we only consider the angle range from $0^{\circ}$ to $45^{\circ}$. For  $\mathrm{Fe_2Br_2}$, the softest/hardest
direction  is along the (100)/(110)
direction with  Young¡¯s moduli of 38.03 $\mathrm{Nm^{-1}}$/64.04 $\mathrm{Nm^{-1}}$.
 The maximum value of $C_{2D}$  is
less than that of graphene (340 $\mathrm{Nm^{-1}}$)\cite{gra}, indicating its extraordinary
flexibilities. This provides possibility to use strain to tune physical properties of $\mathrm{Fe_2Br_2}$.
The minima/maxima of $\nu_{2D}(\theta)$  is 0.104/0.468 along the
(110)/(100) direction.

\begin{figure}
  \includegraphics[width=8cm]{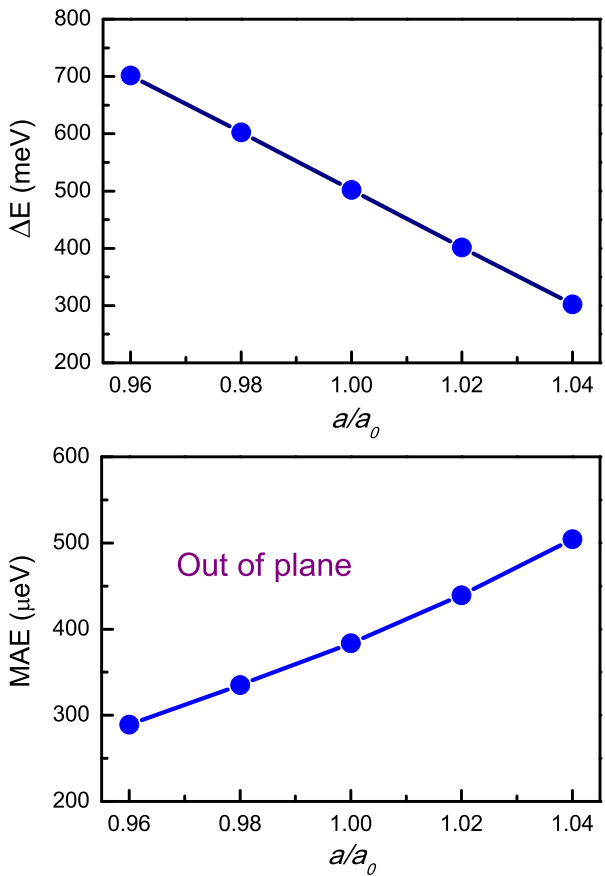}
\caption{(Color online)For monolayer $\mathrm{Fe_2Br_2}$, the energy difference between AFM and FM (Top) and MAE (Bottom) as a function of  $a/a_0$ at $U$=2.50 eV. }\label{s-em}
\end{figure}

Without  SOC, $\mathrm{Fe_2Br_2}$ is a 2D half Dirac semimetal state with a large-gap insulator for spin up
and a gapless Dirac semimetal for spin down. The band crossings are protected by $M_y$ ($M_x$),  forbidding  $d_{z^2}$ and $d_{xy}$ to hybridize (see FIG.4 of ESI).
In contrast to typical
Dirac cones in graphene, there are  four Dirac cones in 2D BZ due to  $C_4$ symmetry. Within SOC, the Dirac gap of of 166 meV can be produced.
These similar results can be found in  $\mathrm{Fe_2I_2}$,  $\mathrm{Fe_2IX}$ (X=Cl and Br), $\mathrm{FeI_{1-x}Br_{x}}$ (x=0.25 and 0.75), $\mathrm{Li_2Fe_2X_2}$ (X=S, Se and Te) and  $\mathrm{Li_2Fe_2SSe}$\cite{fe,fe1,fe2,fe3,fe4}.
The magnetic anisotropy direction is very important to determine the topological properties of some 2D systems.
For example, for monolayer $\mathrm{VSi_2P_4}$, $\mathrm{FeCl_2}$ and FeClF\cite{h10,h11,h12}, the QAH states can exist at proper $U$ range with out-of-plane magnetic anisotropy, and no special QAH states appear with in-plane case. The  energy band structures  with out-of-plane and in-plane magnetizations by using GGA+SOC are plotted in
\autoref{band-2.5}. By
varying the magnetization orientation from $z$ to $x$, the band gap  decreases down to very small value (Strictly, the gap should be zero.).
The topological
edge states with out-of-plane and in-plane magnetizations are calculated, as shown in \autoref{s-2}. For in-plane case, no non-trivial chiral edge modes appear within the bulk gap of $\mathrm{Fe_2Br_2}$, meaning its Chern number $C$=0.

To corroborate this finding, the Chern number is recalculated  by integrating the Berry curvature ($\Omega_z(k)$) of the
occupied bands:
 \begin{equation}\label{pe1-4}
  C=\frac{1}{2\pi}\int_{BZ}d^2k \Omega_z(k)
\end{equation}
\begin{equation}\label{pe1-4}
  \Omega_z(k)=\nabla_k\times i\langle\mu_{n,k}|\nabla_k\mu_{n,k}\rangle
\end{equation}
in which  $\mu_{n,k}$ is the lattice periodic part of the Bloch wave functions.
The distributions of Berry curvature in 2D BZ with out-of-plane and in-plane magnetization are plotted in \autoref{s-2}.
For out-of-plane magnetization,  the
hot spots in the Berry curvature  are around four gapped Dirac cone, which have the same
signs. A quantized
Berry phase of $\pi$ for each gapped Dirac cone can be attained, and the total Berry phase of 4$\pi$  due to four Dirac cones  means a
high Chern number $C$=2. For in-plane magnetization,  there are four main
hot spots in the Berry curvature along four $\Gamma$-M lines, and two of  then  have the opposite
signs with the other two, giving rise to disappeared Chern number.
\begin{figure}
  \includegraphics[width=7cm]{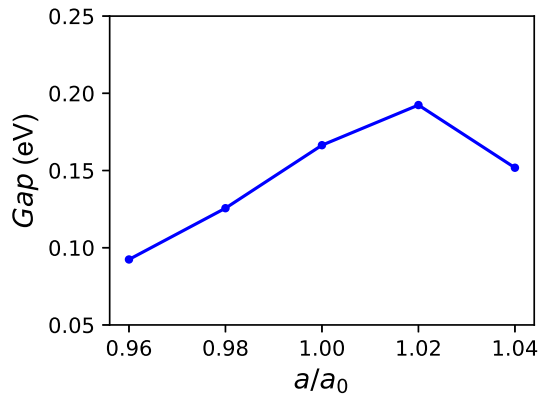}
\caption{(Color online)For monolayer $\mathrm{Fe_2Br_2}$, the gaps as a function of  $a/a_0$ at $U$=2.50 eV. }\label{s-gap}
\end{figure}

\begin{figure*}
   \includegraphics[width=16cm]{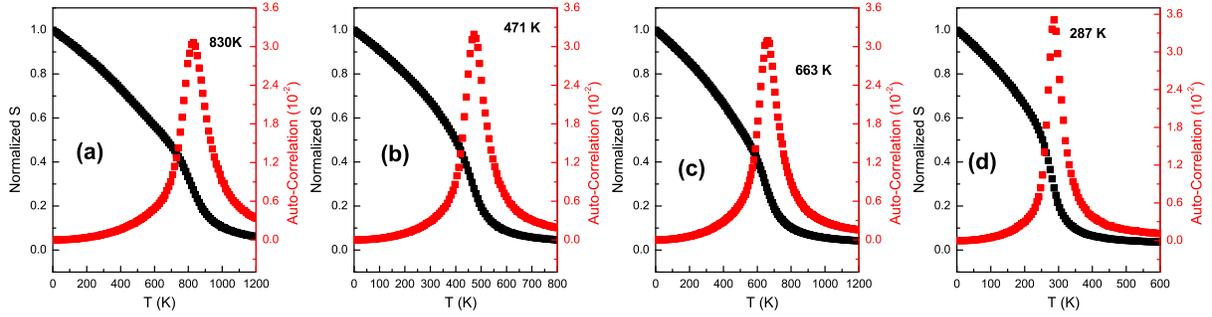}
  \caption{(Color online)For $\mathrm{Fe_2Br_2}$ monolayer, the normalized magnetic moment (S) and auto-correlation  as a function of temperature with $U$$=$1.50 eV (a), 2.50 eV (b) and $a/a_0$=0.96 (c), 1.04 (d) at $U$$=$2.50 eV.}\label{tc}
\end{figure*}

For $\mathrm{Fe_2Br_2}$, a large QAH gap is not because  the effective
SOC strength of $d$ orbitals is enlarged by bonding with
heavy elements Br, and is because  the electron correlation improves the SOC
effect. To illustrate this, the  energy band structures with  only considering SOC of Fe/Br with out-of-plane magnetization are plotted in \autoref{band-2.5}. For only considering SOC of Fe, the gap is 140 meV, which is very close to 166 meV with full SOC.
However, the gap of 35 meV is attained with only considering SOC of Br, which is about one-fifth of one with full SOC.
So, the large gap is mainly from the SOC effects of Fe combined with electronic correlation.

To explain the key role of SOC of Fe, the atom- and Fe-3$d$-orbital-resolved band
structures of $\mathrm{Fe_2Br_2}$   are plotted in FIG.4 of ESI by using GGA+SOC.  It is found that the Fe-$d$ derived states contribute mainly to the gapped Dirac states near the Fermi level, which leads to importance of SOC of Fe to induce QAH gap.  Without SOC, the
two crossed spin-down bands (see \autoref{band}) near the Fermi level are mainly
contributed by $d_{z^2}$ and $d_{xy}$ orbitals of Fe, and  their band order is inverted between $\Gamma$ and X.
When considering SOC,  a Dirac gap is produced. Topological
edge states of $\mathrm{Fe_2Br_2}$ monolayer  with only considering SOC of Fe/Br are plotted in FIG.5 of ESI.
Two chiral gapless edge modes appear within the bulk gap for only considering SOC of both  Fe and Br, indicating QAH properties.

\section{Strain effects}
Strain can  modify the distance between atoms, and then can tune kinetic and interaction energies of electrons, which can induce topological transition\cite{v6,v7,v8}.
The biaxial  strain effects on QAH robustness of  $\mathrm{Fe_2Br_2}$  are investigated, and $a/a_0$ (0.96-1.04) is used to simulate the biaxial strain with   $a$/$a_0$ being the strained/unstrained lattice constants. The $a/a_0$$<$1/$>$1 means compressive/tensile strain. Based on energy difference between AFM and FM (see \autoref{s-em}),
the ground state of $\mathrm{Fe_2Br_2}$ is
FM ordering  in considered strain range. It is found that compressive strain can enhance FM interaction, and tensile strain can weaken one, which will produce important effects on Curie temperature of  $\mathrm{Fe_2Br_2}$. Based on previous research results,  the magnetic orientation is very key to induce QAH state, and the MAE as a function of $a/a_0$ is plotted in  \autoref{s-em}. In considered strain range, the  out-of-plane magnetic anisotropy is robust against strain, which confirms  possible existence of QAH state.

The GGA+SOC energy band structures  of $\mathrm{Fe_2IBr}$ vs $a/a_0$ are plotted in FIG.6 of ESI, which show that they are all FM semiconductors.
The gaps as a function of strain are shown in \autoref{s-gap}, which shows a nonmonotonic behavior
with $a/a_0$ from 0.96 to 1.04. With increasing $a/a_0$, the gap firstly increases, and then decreases, which is due to the change of conduction band bottom from one point along $\Gamma$-X to $\Gamma$ point.
The topological
edge states of strained  monolayer $\mathrm{Fe_2Br_2}$  are  calculated, and  two chiral topologically
protected gapless edge states emerge at representative 0.94 and 1.06 strains (see FIG.7 of ESI), giving  Chern number C=2.
These prove  that  the  QAH state of  monolayer $\mathrm{Fe_2Br_2}$  is robust against strain. Unlike $\mathrm{Fe_2Br_2}$,  strain can induce a series of phase transitions  from ferrovalley (FV) insulator
to  half-valley metal (HVM) to QAHI in these 2D materials\cite{v6,v7,v8}, for example $\mathrm{VSi_2N_4}$.

\section{Curie temperature}
Based on the previous discussion, both electronic correlation and strain can effectively tune FM interaction of $\mathrm{Fe_2Br_2}$, which means that they can affect observably its Curie temperature ($T_C$). The MC simulations
 are performed to estimate  $T_C$ of monolayer $\mathrm{Fe_2Br_2}$ based on spin
  Heisenberg model,  whose Hamiltonian can be expressed as:
  \begin{equation}\label{pe0-1-1}
H=-J\sum_{i,j}S_i\cdot S_j-A\sum_i(S_i^z)^2
 \end{equation}
where  $S_i$/$S_j$, $S_i^z$,  $J$ and  $A$   are   the
spin vector of each Fe atom, spin component parallel to the $z$ direction,  nearest neighbor exchange parameter and   MAE.

Based on the energy
difference between AFM and FM,   the normalized
  magnetic coupling parameters ($|S|$ = 1)
can be attained as $J$=($E_{AFM}$-$E_{FM}$)/8.
At representative correlation strength ($U$=1.5 eV and 2.5 eV) and strain ($a/a_0$=0.96 and 1.04),
the  calculated $J$ values are 110.2 meV, 62.8 meV,  87.7 meV and 37.8 meV, respectively. We plot
 the normalized magnetic moment and auto-correlation  as a function of temperature  in \autoref{tc}.
The estimated  $T_C$ is about 830/471 K for $U$$=$1.5/2.5 eV, and 663/287 K for $a/a_0$$=$0.96/1.04. So, both electronic correlation and strain have important influence on  $T_C$ of $\mathrm{Fe_2Br_2}$.  It is found that the reduced correlation strength and compressive strain can improve $T_C$.
At typical $U$=2.5 eV, the predicted $T_C$ is significantly higher than that of previously
reported many  2D FM semiconductors (20-160 K)\cite{m7-6,tc1,tc2}. These indicate that $\mathrm{Fe_2Br_2}$ should be a room-temperature QAHI.

\section{Discussion and Conclusion}
Electronic correlations have significant effects  on physical  properties of materials with localized $d$ electrons, especially for low-dimensional systems. For example monolayer $\mathrm{VSi_2P_4}$, $\mathrm{FeCl_2}$ and FeClF\cite{h10,h11,h12},
 different correlation strengths can drive these systems into various types of interesting ground states,  such as FV, HVM and QAH states. The correlation strength can also tune magnetic anisotropy of these 2D materials. For monolayer $\mathrm{VSi_2P_4}$, several transitions in the magnetic anisotropy can be observed with varied $U$\cite{h10}. The  out-of-plane FM ordering  allows a nonvanishing Chern
number of these 2D systems\cite{h10,h12}, which is different from in-plane FM situation. Here, the correlation can be used to induce enhanced SOC effect of Fe atoms in $\mathrm{Fe_2Br_2}$, which can produce large-gap QAHI with out-of-plane magnetic anisotropy. In fact, amplifying the SOC effect in light elements can be achieved by the interplay between particular crystal symmetry and electron correlation, which contains certain partially occupied orbital multiplets\cite{h13}. In a word, our works emphasize the importance of electronic correlation and magnetic anisotropy to determine the electronic state of $\mathrm{Fe_2Br_2}$.

In summary,  the  intriguing large-gap of QAHI $\mathrm{Fe_2Br_2}$ is  explained by the reliable first-principle calculations, which is due to
correlation-enhanced SOC effects of Fe element. The $\mathrm{Fe_2Br_2}$  exhibits excellent dynamic, mechanical and thermal stabilities, and can realize room-temperature high-Chern-number ($C$=2) QAHE. At representative $U$=2.5 eV\cite{fe,fe5},  the band gap of 166 meV and $T_C$ of
about 471 K are obtained, respectively. The emergence of
 QAH states in monolayer $\mathrm{Fe_2Br_2}$ is robust against strain. The $T_C$ can be enhanced
to 663 K at 0.96 compressive strain, and  the band gap remains at about 92 meV. At 1.04 tensile strain, the $T_C$ of 287 is still close to room temperature, and the gap is about 152 meV.   Our works provide a comprehensive understanding  of correlation effects on large nontrivial band gap in  monolayer  $\mathrm{Fe_2Br_2}$, which can be readily extended to other monolayer $\mathrm{Fe_2XY}$ (X/Y=Cl, Br and I). These results are also helpful to deepen our understanding of large-gap QAHI in light-element materials.

\begin{acknowledgments}
This work is supported by Natural Science Basis Research Plan in Shaanxi Province of China  (2021JM-456),  the Nature Science Foundation of China (Grant No.11974393) and the Strategic Priority Research Program of the Chinese Academy of Sciences (Grant No. XDB33020100). We are grateful to the Advanced Analysis and Computation Center of China University of Mining and Technology (CUMT) for the award of CPU hours and WIEN2k/VASP software to accomplish this work.
\end{acknowledgments}

\end{document}